# Label-free single nanoparticle identification and characterization including infectious emergent virus

**Minh-Chau Nguyen [1], Peter Bonnaud [1], Rayane Dibsy [2], Guillaume Maucort [3], Sébastien Lyonnais [4], Delphine Muriaux [2,4], Pierre Bon [1,*]**

1 – Université de Limoges, CNRS, XLIM, UMR 7252, F-87000 Limoges, France

2 – IRIM (Institut de Recherche en Infectiologie de Montpellier), Université de Montpellier, UMR 9004 CNRS, Montpellier, France

3 - Laboratoire Photonique Numérique et Nanosciences, University of Bordeaux, F-33400 Talence, France; LP2N UMR 5298, Institut d'Optique Graduate School, CNRS, F-33400 Talence, France

4 – CEMIPAI, Université de Montpellier, UAR 3725 CNRS, Montpellier, France

*Corresponding author: pierre.bon@cnrs.fr*

**Screening of unknown particles, including viruses and nanoparticles, is key in medicine, industry and pollutant determination. However, existing techniques require sample *a priori* knowledge or modification (e.g. fluorescence). Here we introduce RYtov MIcroscopy for Nanoparticles Identification (RYMINI), a non-invasive and non-destructive optical approach that is combining holographic label-free 3D tracking and high-sensitivity quantitative phase imaging into a compact optical setup. Dedicated to the characterization of nano-objects in solution, it is compatible with highly demanding environments such as level-3 biological laboratories. Metrological characterization has been performed at the level of each single object on both absorbing and transparent particles as well as on infectious HIV-1, SARS-CoV-2 and extracellular vesicles in solution. We demonstrate the capability of RYMINI to determine the nature, concentration, size, complex**

**refractive index and mass of each single particle. We discuss the application of the method in unknown solution without requiring any knowledge or model of the particles' response. It paves the way to label-free nano-object identification in *terra incognita*.**

## *Main*

Having access to metrological parameters on individual nanoparticles (NPs) is key to characterize the production for food, industry and biomedicine [1–3] and to identify relevant differences of nature between particles[4]. In this quest, some measurements could unlock unprecedented comprehension on each individual nano-object including the molecular content and the density of matter which involve a knowledge on both dimension and weight of each particle in their native environment. Ensemble measurements using dynamic light scattering (DLS) have been proposed [5,6] to get access to the size of particles –even in polydispersed solution- but the intrinsic lack of specificity in the measured signals make this method challenging when nanoparticles with identical size distribution but different nature coexist in the same solution (e.g. empty and full extracellular vesicles).

At the single particle level, different techniques have been proposed to measure the object size without sample modification (e.g. fluorescence labeling). Based on light scattering [7] or single-particle tracking (SPT) [8–15] they can furthermore extract a signal related to the particle refractive index mismatch with the solvent. *A priori* knowledge about the particle light/matter interaction is then required to sort the particles, making these methods very challenging when applied to both absorbing and transparent nano-objects. Moreover, when working with nanometer-size bio-objects such as vesicles or viruses, having access to the particle refractive index is not enough to perform acute biological interpretations and the development of the technique remains confidential outside of the optical community. At microscopic scale, it has been demonstrated in the 1950s[16] that light-matter interaction can be converted into mass measurement for biological samples. This approach has been widely applied on optically resolved samples including eukaryotic cells[17,18] and bacteria[19]. Different methods of NP or virus mass measurement with limited *a priori* knowledge have been introduced by using electron microscopy[20,21], mass spectroscopy[22,23], nanometer-scale pores[24]. However, these methods are only applicable either as destructive method, or for fixed particles.

We introduce **RYtov MIcroscopy for Nanoparticles Identification (RYMINI)** as a novel tool to identify each nanoparticle in solution and characterize its size, concentration, optical attenuation and molecular mass without *a priori* particle optical response knowledge. It unlocks the capability to detect and characterize unknown objects including emergent virus or pollutant. We have combined the capability of holography for (i) imaging in a single 2D image large 3D volume and for (ii) numerical refocusing of each freely diffusing particle[25,26], with the high-stability, compactness and sensitivity of self-reference interferometry[27,28]. This grants an access to the size *d* of each particle (*via* SPT analysis), its complex refractive index $ñ = n + i \cdot k$ (and thus its electric susceptibility $\chi = ñ^2 - 1$) and its mass *m* (in Dalton or gram). Moreover, the imaging volume, a key element to measure particle concentrations in single-shot, is also precisely determined *a posteriori*. Since the setup aims at the characterization of infectious viruses, it is designed to be fully-compatible with acquisitions in demanding environments, including biosafety cabinets.

## Acquisition and processing method

The core idea of our method is to simultaneously measure the size and the optical intensity & phase responses of each nano-object. These three independent measurements grant signal specificity to determine key metrological information and thus to identify each nanoparticle. To do so, we image freely diffusing nanoparticles (following Brownian motion) in their native solution with an optical microscope. The illumination has a controlled spatial and temporal coherence (super-continuum laser, see methods and SI.2) and each particle –even far from the imaging plane– forms an image on the sensor (Fig 1.a). By using a detector sensitive to both the phase $\varphi$ and the intensity $I$ of the light, it is possible to numerically compute a 3D image stack from a single 2D hologram. The only requirement is the knowledge of the illumination wavelength $\lambda$ and the microscope parameters (pixel size in the object space and numerical aperture). In order to keep our localization method independent from the particle nature (*i.e.* absorbing, transparent or in-between), we have merged the intensity $I$ and the phase signal $\varphi$ into a novel observable, called Rytov intensity:

$$I_{Rytov} = |E_{Rytov}|^2 = \left| \frac{i\,\lambda \cdot n_m}{\pi} \left[ \frac{\ln(I)}{2} + i \cdot \varphi \right] \right|^2,$$

with $E_{Rytov}$ the Rytov field (generalized from the polarizability tensor[29]), and $n_m$ the surrounding medium refractive index. The real part (respectively the imaginary part) of the Rytov field is linked to the real part (respectively the imaginary part) of the refractive index difference between the particle and the medium. Even though obtained with a label-free technique, the Rytov intensity image looks very similar to single fluorescent particle image (see Fig.1.b). Regular and robust localization algorithm can therefore be applied to 3D super-localize the particle (Gaussian fitting, see methods) without requiring *a priori* knowledge about the sample nature. The localization precision is far beyond the microscope resolution (e.g. λ/10 for SARS-CoV-2 virus) within imaging volume of typically 20-pL (~28×28×25 µm3, see Fig.2.k and SI.6 for complete characterization). This allows unbiased particle concentration measurements by counting the number of detected objects in this numerically tunable reconstruction volume. Successive temporal holograms are then recorded and both the 3D position over time and the Rytov field image of each particle at each time point are extracted. Though the particles are moving, their images can be numerically refocused and remains static over time at their best focus plane. Hence, each particle image can be averaged through time to obtain a maximized signal-to-noise ratio (SNR) intensity and phase image of each freely diffusing particles extracted (see Movie SI.1). We call this approach **Dynamic Average Imaging (DAI)**. After DAI, the image SNR increases, theoretically by $\sqrt{N}$ factor (N being the number of tracking frames), which largely improves the characterization of particles.

Since the nanoparticles are diffusing following a Brownian motion, according to the Stokes-Einstein law[30], the particle diameter *d* can be obtained *via*:

$$d = \frac{k_B T}{3\pi \eta D} ,$$

with $k_B$ is the Boltzmann constant, *T* the fluid temperature, *η* the viscosity of the fluid and *D* the diffusion coefficient. The classical way to measure *D* is to compute the mean squared displacement (MSD) of each nanoparticle. Our holographic method gives access to the 3D position and thus the mean squared displacement (MSD) is more precise than other methods working in 2D with equivalent localization precision[31]. The tracking has been performed at high frame rate (400-Hz) to ensure accurate evaluation of diffusion coefficient *via* MSD and to minimize the contribution from environment instabilities (thermal, vibrations…). We were able to perform measurements even inside a biosafety cabinet of confined level-3 biological laboratory (see Fig. 3.e). To increase size evaluation precision from MSD, we have used a unique characteristic of our method: the Rytov field depends jointly on particle nature and size which allows us to perform correlation between particles which exhibit the similar Rytov field (see SI.4). Moreover, our approach unlocks concentration determination without *a priori* knowledge on the particle type: the numerically reconstructed imaging volume is automatically adjusted by the signal-to-noise ratio of the detected particles.

Noteworthy, the independent measurements of both the diameter $d$ and the Rytov field $E_{Rytov}$ integrated over the whole particle image give access to useful bio-physical parameters for each NP. This includes the NP mass $m$ (in gram or Dalton) and its complex refractive index $ñ = n + i \cdot n'$ following:

$$m = -\frac{\rho_{obj}}{2\, n_m\, (n - \text{Re}(n_m))} \iint_{x,y} \text{Re}(E_{Rytov}) \mathrm{d}x\mathrm{d}y$$

$$= -\frac{\lambda\, \text{Re}(n_m)\, \rho_{obj}}{2\pi(n - \text{Re}(n_m))} \iint_{x,y} \varphi \cdot \mathrm{d}x\mathrm{d}y = -\beta \iint_{x,y} \varphi \cdot \mathrm{d}x\mathrm{d}y,$$

$$ñ = \sqrt{1 + \chi} = n_m + \frac{3\,\lambda\, \text{Re}(n_m)}{\pi^2 d^3} \iint_{x,y} \left(-\varphi + i\,\frac{\ln(I)}{2}\right) \mathrm{d}x\mathrm{d}y,$$

with $n_m$ the medium (complex) refractive index, $\lambda$ the imaging wavelength and $\rho_{obj}$ the bulk material density (see SI.1 for complete discussion). **Figure 1.b** and Movie SI.1 give an overview of the image processing and information harvesting.

## Results - Discussions

**Characterization of nanoparticle solutions**

First we have studied with RYMINI monodispersed solutions containing dielectric (polystyrene, silica, diamond) or plasmonic (gold, silver) particles. Their statistics are summarized in the **Table 1**.

In Fig.2.a-c, a solution of 100-nm polystyrene (PS) nanoparticles has been imaged and both the complex refractive index and the size of each particle are presented. The NP diameter is measured to be $d = 105 \pm 13$ nm which agrees with the information provided by the manufacturer, in which the average size is $102 \pm 3$ nm, with a dispersion of 7.6 nm. PS nanoparticles with size ranging from 60-nm to 200-nm have been studied with the same approach (Fig 2.d,e and Movie SI.2) showing stability in the measured values for both refractive index and density. The median complex refractive index has been measured $\tilde{n} = \left(1.543^{+0.094}_{-0.063}\right) + i \cdot \left(0.005^{+0.081}_{-0.086}\right)$ which is close to the PS bulk refractive index at $\lambda = 450 nm$ [32], $\tilde{n}_{PS} = 1.6104 + i \cdot 6.33 \cdot 10^{-7}$, confirming no plasmonic effect on these particles. One can notice that the analyzed batch of 100-nm PS nanoparticle exhibits a statistically relevant difference in density indicating a possible difference in fine composition. In addition to PS particles, Silica (Si) and diamond (C) di-electric nanoparticles as well as metallic nanoparticle including gold (Au) and silver (Ag) were studied in Fig.2.f-i and Movie SI.3.

For plasmonic particles, the light-matter interaction is very dependent on the particle size and nature as well as the working wavelength[29]. Considering the metal particles studied in Fig.2.f-i (60-nm and 100-nm Au NP, and 100-nm Ag NP), they exhibit a plasmon resonance when illuminated at respectively 530-nm, 570-nm and 480-nm, according to Mie theory[33]. The module of the complex refractive index is maximized when the illumination is at plasmon resonance wavelength and is approaching the refractive index of the bulk material when the illumination is far from resonances. For two types of particles, Au and Ag, of the same size (100-nm), observed, the measured refractive index are respectively $ñ_{Au_{100}} = \left(1.333^{+0.159}_{-0.101}\right) + i \cdot \left(2.014^{+0.799}_{-0.491}\right)$ and $ñ_{Ag_{100}} = \left(1.338^{+0.345}_{-0.268}\right) + i \cdot \left(4.748^{+1.950}_{-1.497}\right)$ with an illumination at 450-nm. This confirms at this illumination wavelength a much higher plasmon effect for Ag particles since the refractive indices of bulk gold and silver are tabulated[34] at $ñ_{Au_{bulk}@450nm} = 1.38 + i \cdot 1.92$ and $ñ_{Ag_{bulk}@450nm} = 0.04 + i \cdot 2.65$.

Particle concentration measurement requires a knowledge on both the imaging volume and the number of objects in this last. Although the 2D field of view (FOV) is fixed by the magnification and the size of the sensor, the depth of the imaging volume is chosen *via* holographic processing. The maximal depth of FOV is thus related to the signal-to-noise ratio of the system, with a good linearity with the Rytov field amplitude, as shown in Fig. 2.k. Therefore, the 3D volume can be tuned and maximized *a posteriori* without any particles' nature knowledge (see SI.6 for discussion of the 3D volume as a function of different particle Rytov Intensity). For bright particles, such as 100-nm Ag NPs, the depth of field can reach up to 50 μm (from -25 to +25 μm). Our unique capability to numerically fix the imaging volume grants access to single-shot and unbiased particle concentration measurements. In Fig.2.j, concentrations have been measured over different dilutions of a 100-nm PS stock solution of $1 \times 10^{11}$ particles/mL (determined with commercial Videodrop, Myriade, France). In a holographic imaging volume of 20-pL (28×28×25 µm$^3$), the concentration is obtained by counting the NP traces in single time sequence acquisition (~300 images acquired at 400-Hz). Concentrations down to $0.5 \times 10^8$ particles/mL, or **≈0.08 pM** (about 1 particle per holographic volume) -with a precision of $0.4 \times 10^8$ particles/mL (0.06 pM)- were measured, in good agreement with the theoretical values, estimated from dilution proportions.

**Virus identification and characterization at the single particle level**

We next challenged the RYMINI to analyze biological nanoparticles in solutions: (i) artificial HIV-1 Gag virus-like-particles (VLP) and VLP enriched in viral RNA (HIV-1 Gag VLP+RNA)[35] (ii) infectious HIV-1 particles (HIV-1 wt) and HIV-1 particles deprived of the surface Envelope Glycoproteins (HIV-1 ΔEnv), (iii) infectious SARS-CoV-2, and (iv) extracellular vesicles (EVs) secreted in cell culture in absence of VLP or viral infection (Fig.3.a). While EVs and VLP can be handled and measured without biohazard, measurements of infectious class-3 viruses such as HIV-1 and SARS-CoV-2 have to be performed in confined biological environments of level-3 security. The acquisitions of infectious samples were thus performed inside a biosafety cabinet of a confined level-3 biological laboratory (CEMIPAI, Montpellier, France) as shown in Fig.3.e (laser-beam safety panels have been removed to take the picture).

For biological samples we have considered as pertinent to measure the size and mass (in Da) of each particle. At first order approximation, virus dry mass can be inferred from the optical phase image and its size (see SI.5). The results for VLP, viruses and EVs are illustrated in Fig.3.b-d and reported in **Table 2**. HIV-1 Gag-VLP size analysis first validates the capabilities of our method, with a measured diameter of $141^{+35}_{-17}$-nm, which is the expected size of HIV-1 Gag-VLPs produced in higher eukaryotic cells [36,37]. HIV-1 Gag VLP particles enriched in viral RNA (HIV-1 Gag VLP+RNA) showed an increased size of $179^{+25}_{-26}$-nm, which correlates with the size modulation of viral RNA on VLP assembly [35]. When moving to native, wt-infectious HIV-1 particles, the measured diameter ($171^{+25}_{-17}$-nm) was higher to that previously reported by CryoEM ($145 \pm 25$-nm)[38]. This is however consistent since the particle size determined by CryoEM are restricted to the outside of the lipid bilayer, therefore omitting the outer shell of the heavily glycosylated Gp160$^{Env}$, which should strongly participate in the hydrodynamic radius measured here. This was indeed confirmed by the measured size ($139^{+23}_{-14}$-nm) of HIV-1 ΔEnv lacking the surface Gp160$^{Env}$ trimers. SARS-CoV-2 size was found around $85.2^{+6.5}_{-9.4}$-nm, which is consistent with sizes reported so far by CryoEM ($91 \pm 11$-nm)[39] and AFM ($89 \pm 19$-nm)[40]. Finally, samples of extracellular vesicles showed a broader size range ($233^{+54}_{-39}$nm), as expected from these heterogeneous family of biological particles[41]. Mass variations between all these biological nanoparticles correlate with their variations in size and content, with HIV-1 Gag VLP + RNA > HIV-1 wt > HIV-1 ΔEnv > HIV-1 Gag VLP > SARS-CoV-2. EVs can content various biological material such as proteins and nucleic acids and showed an important heterogeneity in mass accordingly. Interestingly, particles effective densities (in Da$^{1/3}$/nm) confirmed that HIV-1 Gag VLP and the EVs are less packed, while HIV-1 VLP Gag + RNA, HIV-1 wt, HIV-1 ΔEnv and SARS-CoV-2 show equivalent densities, probably reflecting viral fitness to fully pack particles in the 80-140 nm diameter range. Altogether, these results demonstrate that RYMINI is a powerful method to assess virus size, mass and densities;

and can be used in high confinement laboratory for rapid and easy infectious virus analysis.

**Automatized identification of NP**

We have developed an architecture based on a convolutional neural network (CNN)[42] for automatized particle classification. The architecture scheme is presented in Fig.4.a and detailed in methods and SI.7. To identify each particle, the algorithm uses as input the complex Rytov image, normalized by the particle volume to obtain signals proportional to the complex refractive index. Since the input has complex values, the CNN classifier is adapted for deep complex layers[43,44].

Figure 4.b illustrates the normalized confusion matrix which presents the algorithm capability to recognize individual nanoparticle nature. Our CNN can precisely classify dielectric, VLP and metallic nanoparticles: about 80% for the majority of dielectric labels and VLP, and 95% for metallic labels. The main errors are attributed to uncleaned abnormal inputs (see SI.7). For viruses, the accuracy depends on their type: it reaches about 86% for SARS-COV-2 but tends to mix up between HIV-1 wt and HIV-1 ΔEnv. This is comprehensible and predictable since HIV-1 ΔEnv contains 2 populations: one minor interpreted as immature particles (with signature in term of size and mass close to HIV-1 wt) and one major considered as mature particles[36]. Extracellular vesicles are also classified with moderate certainty, as expected for such heterogeneous class[45].

## *Conclusions*

Our full optical RYMINI method unlocks for the first time to our knowledge identification and metrological characterization of any nanoparticle solution at the single particle level without requiring *a priori* information on the sample nature. It grants access to the type of each detected particle (including viral objects) as well as single-shot concentration, size, complex refractive index and mass (in gram and Dalton). We have demonstrated its capability in monodisperse solutions of metallic and dielectric NP as well as on viral non-purified solutions directly inside a biosafety cabinet of a level-3 confinement laboratory (BSL3). The sample preparation as well as the acquisition are straightforward since large volumetric imaging is obtained in a single-shot manner with the RYMINI technique. Probing sub-pM concentration solutions requires less than one second and only 10μL of solution. Our current sensitivity is compatible with identification and complete characterization of viral species as small as SARS-COV-2 (~88 nm), or NP as small as 40-nm for metallic particles and 60-nm for dielectric particles (PS). The detection sensitivity could be pushed further by a change in the illumination scheme at the price of reducing the accessible imaging volume. Moreover, the label-free intrinsic signature of each particle can be used to quantify the homogeneity of particles in a non-invasive and biosafety-secured manner. The deployment of machine learning tools unlocks automatized identification of each detected particles. We believe that this method could be a key tool for routine pollution analysis, easy and fast biological risk determination, NP metrology during production as well as fundamental biological and non-biological particle studies.

## *Methods*

**Optical setup.** The setup is based on a home-made microscope using a quadriwave lateral shearing interferometer[27,46] to measure the phase and the intensity of the beam. The sample is illuminated by a supercontinuum fibre laser (Leukos, France) filtered by a short-pass 700-nm dichroic and a $450 \pm 10$ nm optical filter (Thorlabs) to reach 10-mW of illumination power. A Köhler illumination has been made using one achromatic lens (75-mm, Thorlabs) and one aspherical lens (50-mm, Thorlabs). The image is formed with an oil-immersion objective Nikon Plan Fluor 100x NA 1.30 and a 400-mm achromat lens for the tube lens (Thorlabs). The total microscope magnification has been measured to be 205×, using a resolution test target (R3L1S4P, Thorlabs). The interferometer is composed of a Modified Hartmann Mask[47], a relay lens and a 2Me$^-$ full-well capacity CMOS camera (Q-2HFW-hm, Adimec). The system is built on 40×50-cm breadboard and has a total height of 50-cm to be easily handled and introduced in a hood or a biosafety cabinet.

**Sample preparation**. Perforated parafilm on a type 1.5 round coverslip plays the role of a 10μL imaging chamber. The parafilm is first heated up to 80°C so that it is slightly melted and sealed to the glass slide. After adding the solution, the chamber is closed by a second glass slide avoiding possible evaporation or leaks during handling.

**Virus preparation.** For HIV-1 and EVs production, 12.5 million of Human embryonic kidney cells (293THEK cell) were seeded in 10 ml of DMEM growth media 1 day before transfection. At 50-70% confluence, cells were transfected with calcium phosphate precipitate method, with 8 μg total quantity of plasmid for either pcDNA3.1 (Mock), pHIV-1Gag[48], pHIV-Psi-viralRNA [49,50], pHIV-1(NL43) and pHIV-1GagΔEnv[36]. Cell culture supernatant containing viral particles was collected 48 hours post transfection. Supernatant was filtered through 0.45 μm and then purified by ultracentrifugation on cushion of 25% sucrose - TNE buffer (10 mM Tris-HCl [pH 7.4], 100 mM NaCl, 1 mM EDTA) at 100000g, for 1 hour 30 minutes at 4°C, in SW32Ti Beckman Coulter rotor. Dry pellet was resuspended in TNE buffer at 4°C overnight.

For SARS-CoV-2 particle production, the strain BetaCoV/France/IDF0372/2020, was supplied by the National Reference Center for Respiratory Viruses hosted by Institut Pasteur (Paris, France) and headed by Pr. Sylvie van der Werf. The human sample from which strain BetaCoV/France/IDF0372/2020 was isolated has been provided by Dr. X. Lescure and Pr. Y. Yazdanpanah from the Bichat Hospital, Paris, France. Moreover, the BetaCoV/France/IDF0372/2020 strain was supplied through the European Virus Archive goes Global (EVAg) platform, a project that has received funding from the European Union's Horizon 2020 research and innovation program under the grant agreement No 653316. SARS-COV-2 was propagated in VeroE6 cells with DMEM containing 2.5% FBS at 37°C with 5% $CO_2$ and harvested 72 hours post inoculation. Virus stocks were stored at -80°C and tittered using plaque assays as previously described[40] and provided by CEMIPAI facility.

**Data acquisition and analysis.** A home-made program based on LabView has been developed for both acquisition and processing. Between 200-300 frames per sample where acquired at 400-Hz with about 1.0 ms integration time (total acquisition time of ≈700-ms) before being analysed for single particle tracking. The analysis algorithm is performed offline with following the workflow: (1) to calculate the phase and intensity from each interferogram (raw camera image); (2) to generate a 3D stack from each intensity/phase couple using numerical propagation; (3) to compute the Rytov intensity of each image in each stack; (4) to pre-detect each particle on a maximum intensity projection along the propagation axis of the Rytov intensity stack; (5) to 3D superlocalize on the Rytov intensity stack each particle from its predetection using 3D gaussian fitting; (6) to extract a sub-pixel register at-the-focus intensity/phase image for each particle using the measured 3D position; (7) to compute the DAI from the registered images and the particle size using the MSD from the temporal particle trace. Traces shorter than 20 images where deleted from the analysed data to ensure a proper size measurement and DAI.

All statistics in the text and tables are in **median and 25/75% percentile values**.

**Machine Learning architecture and hyperparameters.** Our CNN training, implemented with Adam optimizer, is composed by two different parts: 4 feature extraction stages followed by 2 fully connected stage. The first two feature extraction stages are composed of a complex convolution layer, complex batch normalization layer and complex ReLU activation[43,44]. The two first feature extraction stage uses 3×3 kernel and no spatial reduction (using a symmetric padding to avoid border effect). The spatial reduction is obtained with the two last feature extraction stage that uses 3×3 kernel and divide by 2 the input spatial dimensions with stride in convolution (convolution kernel is applied each 2 pixels instead of each pixel). The learning rate is initially set at 0.01 and then is reduced by a factor of 0.95 after each 3 training epochs in order to improve the training convergence. The model was trained during 50 epochs with a batch size of 200 Rytov images of $N = 50 \times 50$ pixels and a cross-entropy loss function. 80% of the initial dataset is attributed to the training set, and the remaining 20% are used to test the algorithm.

**Monodispersed NP solution preparation.** Multiple monodispersed solutions have been used in the experiment: regular polystyrene of 60, 80, 100 and 150-nm (Thermofisher), and calibration standard polystyrene of 100 and 200 nm (3K/4K Series Particle Counter Standards, Thermofisher Duke Standards), silica (Sigma Aldrich), Gold NP (Sigma Aldrich), Silver NP (Alfa Aesar), Diamond (Sigma Aldrich). When required, dilutions were performed with de-ionized water (18 MΩ m$^{-1}$).

## *Acknowledgment*

This work was supported by CNRS. Part of this project was funded by the French National Agency against AIDS and Hepatitis (ANRS). RD is a recipient of a SIDACTION fellowship. This project has received funding from the European Research Council (ERC) under the European Union's Horizon 2020 research and innovation programme (grant agreement No. [848645]).

## *Author contributions*

MCN and PBon designed the experiment, wrote the acquisition and analysis code. MCN built the optical setup and performed the acquisitions and analysis. PBonnaud and GM wrote the deep learning algorithm. RD performed cell culture, infection, sample preparation of viruses. SL supervised setting up the material in BSL3; SL, DM, MCN and PBon performed the experiments in the BSL3. MCN, SL, DM and PBon wrote the manuscript. DM and PBon raised fundings. All authors discussed the data and agreed on the final manuscript.

## *Competing interests*

MCN salary has been financed by the French company *Myriade* between 01/2020 and 05/2021.

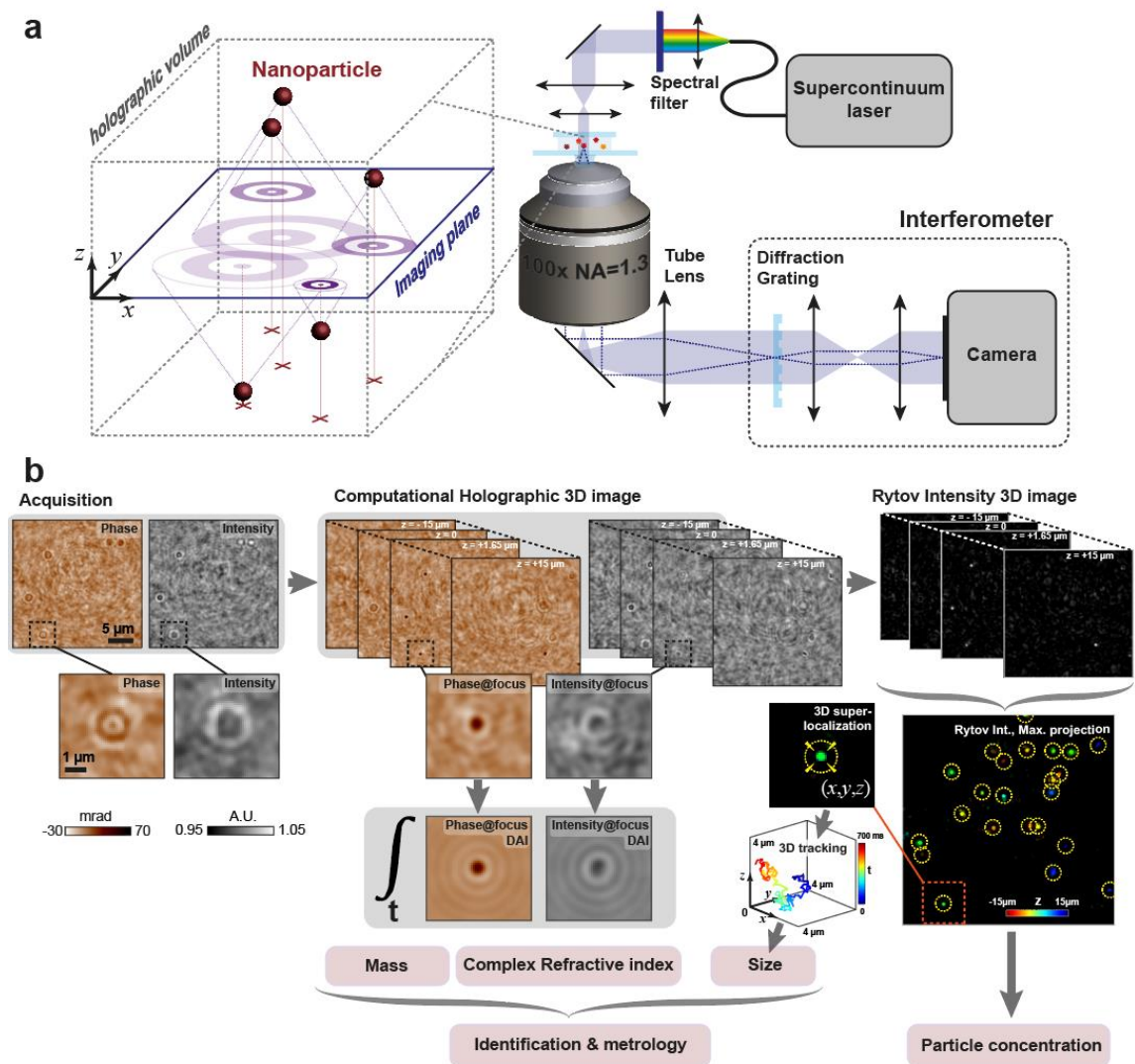

**Figure 1: Principle of experiment.** **(a)** Experimental high-sensitivity holographic setup. **(b)** Analysis workflow. From one intensity/phase image couple acquired in single shot, a 3D image stack is computationally generated. The Rytov field intensity is calculated to unlock each particle localization in 3D. By repeating the procedure at different time points, single particle tracking and signal averaging can be performed to identify and characterize each particle in the solution. DAI stands for Dynamic Average Image (enhancement of the image SNR by temporal averaging of refocused images).

top4topPage 24 of 28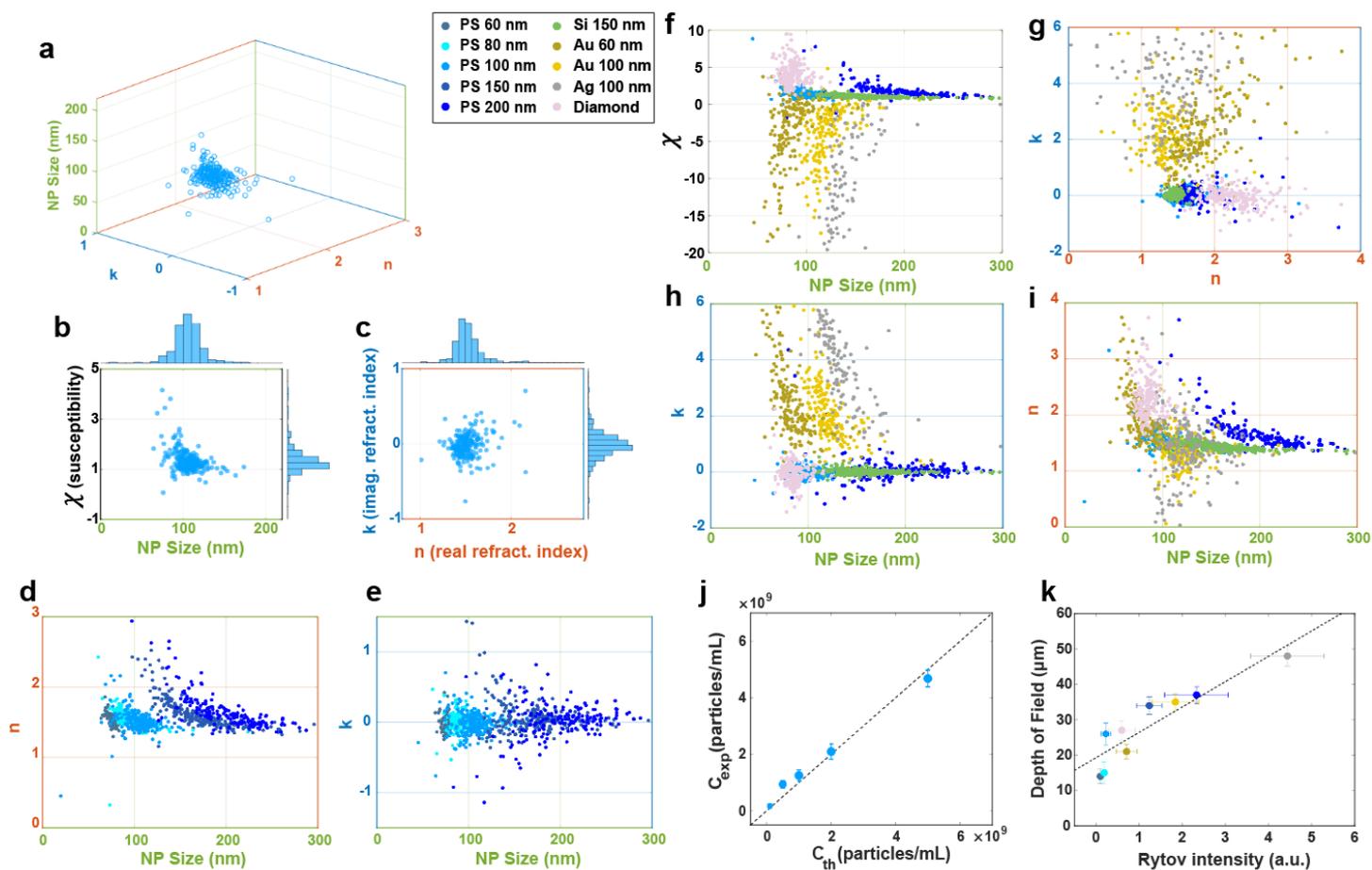

**Figure 2: Quantification of individual NP in solution. (a, b, c)** Analysis of 100-nm PS solution reveals size, complex refractive index, and electric susceptibility. It is illustrated either in 3D (a) or in two 2D graphs (b, c). **(d, e)** PS nanoparticles of different sizes with measurements of real and imaginary refractive indices. **(f, g, h, i)** Measurements for different types of nanoparticles: dielectric (PS, Silica, Diamond) and metal NP (Au, Ag). Their response clusters are separated in the graphs of their optical properties and size, enabling the ability NP's nature identification. (**j**) Single-shot concentration measurement ($C_{exp}$) of 100-nm PS solutions -made from dilutions of an initial solution- compared to the theoretical concentrations ($C_{th}$). (**k**) Depth of field determination as a function of Rytov intensities for different types of particles.

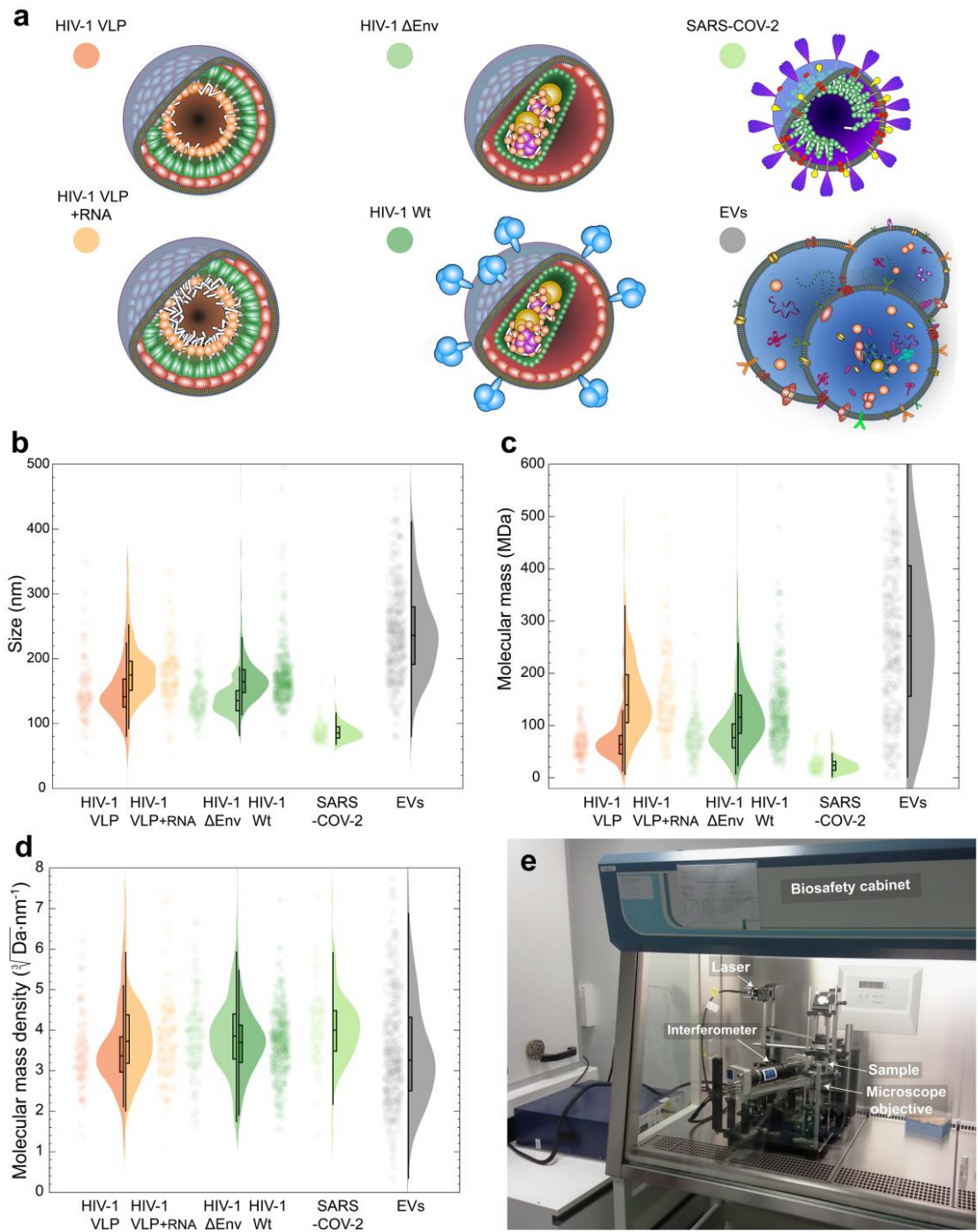

**Figure 3: Quantification of biophysical parameters of viruses and organic nano-objects. (a)** Scheme of the different viruses and virus-like-particles (VLP) analyzed in the experiment. **(b, c, d)** Characterization of biological samples: size, molecular dry mass (MDa) and effective dry mass density (kDa$^{1/3}\cdot$ nm$^{-1}$). **(e)** Picture of the actual RYMINI

microscope siting in a biosafety cabinet in a BSL3 laboratory (CEMIPAI CNRS Montpellier).

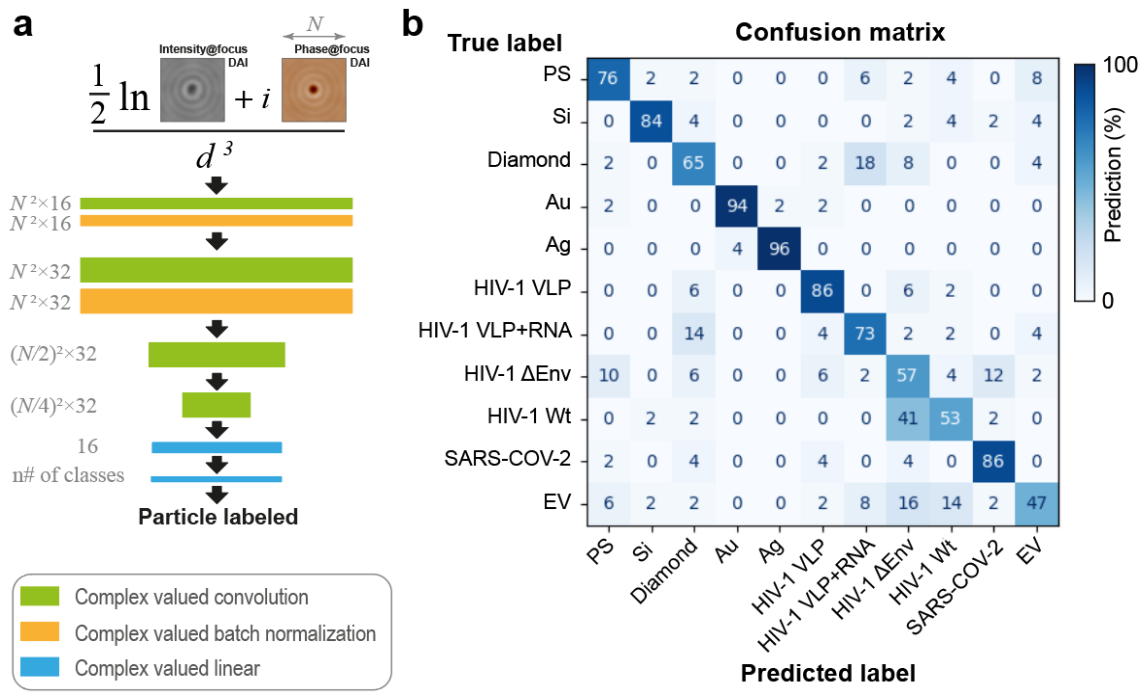

**Figure 4: Particles classification by complex valued convolutional neural network.** **(a)** Architecture scheme with convolutional layers. **(b)** Confusion matrix that represents the classification accuracy after training (in %).

| Particles nature | Number of particles | Hydrodynamic Size (nm) | Refractive index n | Absorption coefficient k | Susceptibility |
|---|---|---|---|---|---|
| PS | 123 | $76.9^{+2.6}_{-2.2}$ | $1.526^{+0.039}_{-0.040}$ | $-0.034^{+0.060}_{-0.055}$ | $1.318^{+0.121}_{-0.125}$ |
|  | 170 | $91.9^{+8.5}_{-6.4}$ | $1.552^{+0.073}_{-0.057}$ | $0.008^{+0.071}_{-0.057}$ | $1.402^{+0.239}_{-0.196}$ |
|  | 221 | $105.1^{+8.2}_{-9.0}$ | $1.504^{+0.072}_{-0.060}$ | $-0.020^{+0.089}_{-0.087}$ | $1.252^{+0.219}_{-0.187}$ |
|  | 228 | $167.8^{+20.5}_{-23.6}$ | $1.568^{+0.140}_{-0.083}$ | $0.025^{+0.089}_{-0.081}$ | $1.443^{+0.413}_{-0.244}$ |
|  | 181 | $200.4^{+28.1}_{-28.0}$ | $1.602^{+0.137}_{-0.103}$ | $0.039^{+0.098}_{-0.121}$ | $1.539^{+0.416}_{-0.327}$ |
| Si | 105 | $124.9^{+15.7}_{-10.2}$ | $1.476^{+0.053}_{-0.046}$ | $0.046^{+0.064}_{-0.058}$ | $1.174^{+0.148}_{-0.140}$ |
|  | 393 | $158.8^{+15.3}_{-13.0}$ | $1.428^{+0.039}_{-0.029}$ | $0.001^{+0.039}_{-0.035}$ | $1.033^{+0.114}_{-0.084}$ |
| Diamond | 208 | $84.6^{+4.0}_{-6.2}$ | $2.241^{+0.223}_{-0.205}$ | $-0.052^{+0.201}_{-0.175}$ | $3.943^{+0.967}_{-0.881}$ |
| Au | 201 | $79.6^{+12.1}_{-7.8}$ | $1.805^{+0.304}_{-0.188}$ | $2.136^{+1.032}_{-0.629}$ | $-2.440^{+1.993}_{-4.529}$ |
|  | 194 | $117.8^{+11.4}_{-7.1}$ | $1.333^{+0.159}_{-0.101}$ | $2.014^{+0.799}_{-0.491}$ | $-3.287^{+2.016}_{-3.702}$ |
| Ag | 191 | $121.9^{+13.5}_{-14.0}$ | $1.338^{+0.345}_{-0.268}$ | $4.748^{+1.950}_{-1.497}$ | $-20.980^{+10.834}_{-22.129}$ |

**Table 1: Number of analysed particles, size, real, imaginary refractive index and electric susceptibility of various nano-particles.**

| Particule nature | Number of particles | Hydrodynamic Size (nm) | Mass (MDa) | $\sqrt[3]{\text{Density}}$ (Da$^{1/3}$.nm$^{-1}$) |
|---|---|---|---|---|
| HIV-1 Gag VLP | 90 | $141^{+35}_{-17}$ | $57^{+15}_{-17}$ | $3.41^{+0.27}_{-0.37}$ |
| HIV-1 Gag VLP + RNA | 162 | $179^{+25}_{-26}$ | $131^{+54}_{-32}$ | $3.59^{+0.44}_{-0.32}$ |
| HIV-1 wt | 337 | $171^{+25}_{-17}$ | $108^{+40}_{-29}$ | $3.48^{+0.38}_{-0.34}$ |
| HIV-1 ΔEnv | 147 | $139^{+23}_{-14}$ | $68^{+24}_{-17}$ | $3.76^{+0.39}_{-0.35}$ |
| SARS-COV-2 | 88 | $85.2^{+6.5}_{-9.4}$ | $18.0^{+5.7}_{-7.4}$ | $3.91^{+0.38}_{-0.63}$ |
| Extracellular Vesicles | 357 | $233^{+54}_{-39}$ | $262^{+130}_{-114}$ | $3.50^{+0.50}_{-0.60}$ |

**Table 2: Size, number of analysed particles, mass and density of various biological nano-objects including viruses.**